\documentclass[pra,a4paper,twocolumn,preprintnumbers,floatfix,aps,nofootinbib]{revtex4}

\usepackage{latexsym}
\usepackage{epsfig}
\usepackage{amssymb}
\newcommand{\bea}{\begin{eqnarray}}
\newcommand{\eea}{\end{eqnarray}}
\newcommand{\be}{\begin{equation}}
\newcommand{\ee}{\end{equation}}

\begin{document}

\title{The matter power spectrum in f(R) gravity}
\date{\today}

\author{Tomi Koivisto}
\email{tomikoiv@pcu.helsinki.fi}
\affiliation{Helsinki Institute of Physics,FIN-00014 Helsinki, Finland}

\begin{abstract}

Modified gravity can be considered as an alternative to dark energy. 
In a generalized theory of gravity, the 
universe may accelerate while containing only baryonic and dark matter. 
We study, in particular, the evolution of matter fluctuations 
in $f(R)$ models within the Palatini approach, and find that the 
resulting matter power spectrum is sensitive to a nonlinear dependence 
on the curvature scalar in the gravitational action. The constraints that 
arise from comparison to the form of the observed matter power spectrum
tighten the previous constraints derived from background expansion
by several orders of magnitude. Models in the allowed parameter space are 
practically indistinguishable from general relativity with 
a cosmological constant when the backround expansion is considered.  

\end{abstract}

\maketitle

\section{Introduction}

One approach to the current dark energy problem in cosmology is to 
consider the observed acceleration of the universe as a consequence of 
gravitational dynamics that deviate from general relativity at 
cosmological scales. 
This can be achieved by modifying the linear proportionality to the 
curvature (Ricci) scalar in the Einstein-Hilbert action to a more general 
dependence on curvature invariants, see\cite{Nojiri:2006ri} for a recent
introduction and\cite{Biswas:2005qr} for other applications of extended gravities 
in cosmology. Here we will consider models in which 
the gravitational Lagrangian is a nonlinear function $f(R)$ of the 
curvature scalar $R$, and the deviations from general relativity  become 
important at small curvature. Such cases have been shown to lead to 
effective dark 
energy\cite{Capozziello:2003tk,Carroll:2003wy, Nojiri:2003ft,Abdalla:2004sw}. 
Furthermore, we adopt the Palatini formulation, in which an $f(R)$ gravity 
is decribed by second order 
field 
equations\cite{Vollick:2003aw,Meng:2003en,Allemandi:2004ca,Allemandi:2005qs}.
The expansion of the universe has been solved in these models and recently also 
succesfully contrasted with various cosmological 
data sets\cite{Capozziello:2004vh,Amarzguioui:2005zq}, see 
also\cite{Sotiriou:2005cd}. Assuming that the correction to the Einstein 
gravity can be parameterized in the form $\sim 
R^\beta$, the data excludes the inverse curvature model (with
$\beta=-1$), but shows a slight preference for a non-null correction,
although the evidence is not compelling\cite{Amarzguioui:2005zq}.

In this paper we will investigate further the viability of the Palatini 
approach to nonlinear gravity by extending the analysis to the constraints
arising from cosmological perturbations.
In a previous work\cite{Koivisto:2005yc} we were concerned that the 
modified matter 
couplings these models feature would, while causing the effective large 
negative background pressure, affect the evolution of inhomogeneities 
in such ways that the primordial fluctuations might not form into 
similar large scale structure as observed at the 
present. We solve cosmology to linear 
order and match the calculated shape of the matter power spectrum 
with the measurements of Sloan Digital Sky Survey 
(SDSS)\cite{Tegmark:2003uf}. The resulting constraints indeed reduce the
allowed parameter space into a tiny region around the simplest 
correction to $f(R)=R$, which is the cosmological constant, corresponding
to $\beta = 0$ and thus $f(R) = R-2\Lambda$. For a recent discussion of 
matter perturbations within the metric formalism of $f(R)$ gravity, 
see\cite{Zhang:2005vt}. 

We begin with a brief review of Palatini approach to 
$f(R)$ gravity and its cosmology in section II. The scheme for deriving the 
large scale structure constraints and the results are presented in section III. 
We conclude in section IV .

\section{Palatini approach to $f(R)$ gravity}

\subsection{Field equations}

We will consider the class of gravity theories represented by the 
action
 \be \label{action_p}
 S = \int d^4x \sqrt{-g}
    \left[\frac{1}{2\kappa}f[R(g_{\mu\nu},\hat{\Gamma}^\alpha_{\beta\gamma})]
    + \mathcal{L}_m(g_{\mu\nu},\Phi,...)\right],
 \ee
where $\kappa = 8\pi G$. The matter action coupling to gravity depends 
only on some matter fields $\Phi,...$ and the metric $g_{\mu\nu}$. Our notation 
emphasizes that the Ricci scalar defining the gravitational sector of the 
action depends on the two independent fields, the metric and the connection 
$\hat{\Gamma}^\alpha_{\beta\gamma}$, see also\cite{Koivisto:2005yc}. 
Explicitly, we may write
 \be \label{ricci_s}
    R \equiv g^{\mu\nu}\hat{R}_{\mu\nu},
 \ee
where
 \be \label{ricci}
 \hat{R}_{\mu\nu} \equiv \hat{\Gamma}^\alpha_{\mu\nu , \alpha}
       - \hat{\Gamma}^\alpha_{\mu\alpha , \nu}
       + \hat{\Gamma}^\alpha_{\alpha\lambda}\hat{\Gamma}^\lambda_{\mu\nu}
        - \hat{\Gamma}^\alpha_{\mu\lambda}\hat{\Gamma}^\lambda_{\alpha\nu}.
 \ee
The field equations which follow from extremization of the action,
Eq.(\ref{action_p}), with respect to metric variations, are
 \be \label{fields2} F \hat{R}^\mu_\nu -\frac{1}{2}f\delta^\mu_\nu = 
      \kappa T^\mu_\nu,
 \ee
where the energy momentum tensor is as usually
 \be \label{memt}
 T_{\mu\nu}^{(m)} \equiv -\frac{2}{\sqrt{-g}} \frac{\delta
 (\sqrt{-g}\mathcal{L}_m)}{\delta(g^{\mu\nu})},
 \ee
and we have defined for notational convenience
\be
F \equiv \partial f/\partial R.
\ee
In general relativity $F = 1$. Contraction of the field 
Eq.(\ref{fields2}) gives 
\be \label{structural}
RF-2f=\kappa T,
\ee
an algebraic relation between the trace of the energy momentum tensor and 
the scalar curvature $R$. 

By varying the action with respect to $\hat{\Gamma}^\alpha_{\beta\gamma}$, 
it is found that the connection is compatible with the conformal metric
 \be \label{conformal}
 \hat{g}_{\mu\nu} \equiv F g_{\mu\nu}.
 \ee
It follows that the Ricci tensor in Eq.(\ref{ricci}) is
 \be \label{riccit}
 \hat{R}_{\mu\nu} = R_{\mu\nu} +
             \frac{3(\nabla_\mu F)(\nabla_\nu F)}{2F^2} -
             \frac{(\nabla_\mu \nabla_\nu F)}{F}
            -\frac{g_{\mu\nu}\Box F}{2F},
 \ee
$R_{\mu\nu}$ being the corresponding tensor associated with the 
metric $g_{\mu\nu}$. Using this we can write the field Eqs.(\ref{fields2}) 
explicitly in terms of the metric $g_{\mu\nu}$. The covariant 
derivative $\nabla$ we refer to is associated with the Levi-Civita 
connection of this metric. Note also that we have assumed that the 
connection is symmetric, whereas in a more general framework of metric 
affine gravity\cite{Puetzfeld:2004yg} torsion is not necessarily absent. 

\subsection{Background cosmology}

We assume a spatially flat FRW universe as the background. There the line 
element is 
\be \label{line}
ds^2 = -dt^2+a^2(t)\delta_{ij}dx^idx^j,
\ee
and matter can be described as a perfect fluid $T^\mu_\nu=diag(\rho,p,p,p)$. 
Since we will consider
late cosmology where $\rho = \rho_m = \rho_c + \rho_b$ ($m$ for matter, $c$ 
for cold dark matter, $b$ for baryons), we assume hereafter that $p=0$ to 
simplify our equations. The modified Friedmann equation is then
 \be \label{friedmann1}
  \left(H+\frac{1}{2}\frac{\dot{F}}{F}\right)^2 = 
  \frac{1}{6F}\left(\kappa\rho + f\right). 
 \ee
An overdot means derivative with respect to the time coordinate $t$ and 
$H$ is the Hubble parameter $H \equiv \dot{a}/a$. 
By using the conservation of 
matter\cite{Koivisto:2005yk} and proceeding as in\cite{Amarzguioui:2005zq}, 
we find that the Hubble 
parameter may be expressed solely in terms of $R$,
\be \label{friedmann2}
H^2 = \frac{1}{6F}\frac{3f-RF}
     {\left[1-\frac{3F'}{2F}\frac{RF-2f}{RF'-F}\right]^2}.
\ee
For a known function $f(R)$, one can be solve $R$ at a given $a$ from 
Eq.(\ref{structural}) and thus get the expansion rate 
Eq.(\ref{friedmann2}) by just algebraic means. In what follows, we will 
also need the derivatives $\dot{H}$ and $\ddot{H}$. Expressions for these 
can be derived similarly\footnote{However, it is unnecessary to report the 
resulting (rather lengthy) formulae here, since when resorting to numerical 
analysis one may equally well evaluate numerically the derivatives of 
Eq.(\ref{friedmann2}), the algebraic formulae providing just a means of a 
consistency check.}.   

In order to investigate these models quantitatively, we must specify the
form the gravitational Lagrangian. We will use the same parameterization 
as\cite{Amarzguioui:2005zq},
\be \label{lagrangian}
f(R) = R - \alpha R^\beta,
\ee
where $\alpha$ is positive and has dimensions of $H^{2-2\beta}$.
The (dimensionless) exponent $\beta$ is less than unity, lest 
the correction to the Einstein-Hilbert action would interfere with the 
early cosmology. Given $\beta$ and the amount of matter in the present 
universe, $\Omega_m \equiv \kappa \rho_m/(3 H_0^2)$, determines the 
scale $\alpha$. Not all combinations of $\alpha$ and $\beta$ 
are consistent with a flat matter dominated universe today, and therefore 
we rather use $\Omega_m$ and $\beta$ as our pair of parameters. 
We will refer to the limit $\beta=0$ as the $\Lambda$CDM case.

\begin{figure}
\begin{center}
\includegraphics[width=0.47\textwidth]{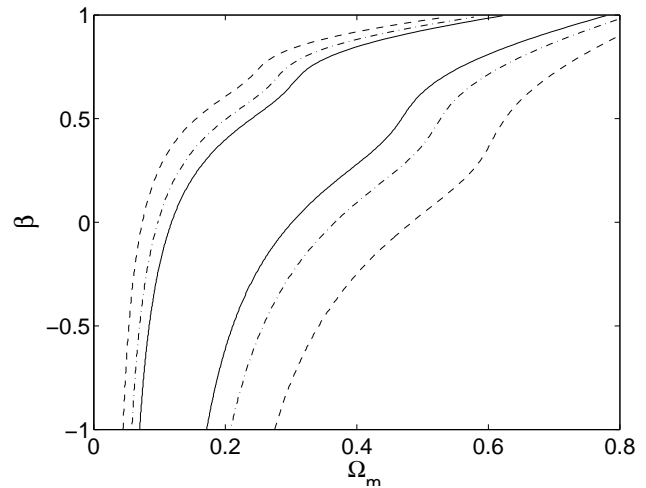}
\caption{\label{shift_limits} The 68, 90 and 99 \% confidence contours 
arising from fitting the CMB shift parameter. 
}    
\end{center}
\end{figure}

To demonstrate that models defined by the Lagrangian (\ref{lagrangian}) 
can produce a plausible expansion history, one can compare the predicted
evolution of the Hubble parameter, Eq.(\ref{friedmann2}) to cosmological 
data. As an example we compute the CMBR (cosmic microwave background radiation) 
shift parameter\cite{Odman:2002wj} given by 
\be
\mathcal{R} = \sqrt{\Omega_m}H_0 \int_0^{z_{dec}} \frac{dz'}{H(z')},
\ee
where $z_{dec}$ is the redshift at decoupling. For these parameters we
use best-fit values found from the CMBR analysis of the WMAP 
team\cite{Spergel:2003cb}, $\mathcal{R} = 1.716 \pm 0.062$ and $z_{dec} = 
1088^{+1}_{-2}$. We plot the constraints resulting from fitting the CMBR 
shift parameter in Fig. \ref{shift_limits}. Projecting to the ($\alpha$,$\beta$) plane 
would reproduce Fig. 1 of\cite{Amarzguioui:2005zq}, where are presented
also further constraints derived from multiple data sets. 

\section{Constraints from the large scale structure}

The foundations of cosmological perturbation theory in the Palatini 
approach to
modified gravity have been laid in\cite{Koivisto:2005yc}. There 
was also derived an evolution equation for the comoving energy density 
perturbation $\delta_m$ in a matter dominated universe. Using the 
time variable $x \equiv \log(a)$, the equation may be written as
\be \label{deltaevol}
\frac{d^2 \delta_m}{dx^2} =  A(x)\frac{d \delta_m}{dx} + B(x) + \frac{k^2}{a^2 H^2}C(x) \delta_m,
\ee   
where the dimensionless functions $A$, $B$ and $C$ are given by
\bea \label{coef_a}
FH^2(2FH + \dot{F})A  & = & - 2\left(\dot{H}+2H^2\right)F^2 \nonumber \\ \nonumber 
 +    2FH\ddot{F} & + & \left(2\dot{F}H+ F\dot{H}-2FH^2\right)\dot{F}, 
\eea
\bea \label{coef_b}
FH^4(  2FH+\dot{F})B   =    2\left(\ddot{H}+2\dot{H}H\right)H^2F^2 & & \nonumber \\ \nonumber 
+   2\dot{H}HF\ddot{F} - \left(2\dot{H}H\dot{F} -  \ddot{H}HF-2\dot{H}^2F
- 2H^2\dot{H}F\right)\dot{F}, & & 
\eea
\bea \label{gradient}
C = -\frac{\dot{F}}{3F(2FH+\dot{F})}.
\eea
We have also checked that our solutions of the Eq.(\ref{deltaevol}) are
consistent with the explicit field equations listed in\cite{Koivisto:2005yc}.

We get the initial conditions for the matter pertubation,
\be
\delta_m = \frac{\rho_c \delta_c + \rho_b \delta_b}{\rho_c+\rho_b},
\ee
by evolving the standard Eistein-Boltzmann equations up until 
$z_i=200$, corresponding to $x_i=-\log(201)$. It is only at smaller redshifts 
that the corrections to 
general relativity begin to have effect. Note also that treating 
the energy density of the universe as consisting of single dust-like 
fluid (made of dark matter and baryons) is well justified. At $z_i=200$ 
radiation is subdominant and may be neglected, and the imprint it has left 
on perturbations in the earlier universe is carried to the initial 
conditions (namely, $\delta_m(k,x_i)$ and $d\delta_m(k,x_i)/dx$). The 
Compton scattering with photons and pressure of the baryons are 
completely negligible at $z_i$ and therefore baryons obey the same 
equations as dark matter. Hence we can, at the linear order, consider 
just the total fluid. 

The effective sound speed\cite{Koivisto:2005nr} is given by 
Eq.(\ref{gradient}), and is nonvanishing except at the $\Lambda$CDM limit. 
The curvature corrections induce effective pressure fluctuations in 
matter, leading to the gradient term in Eq.(\ref{deltaevol}). While the 
other deviations from the perturbation evolution in the $\Lambda$CDM 
scenario may be small when $|\beta|$ is, the gradient can still be large 
at small enough scales. As one could expect\cite{Koivisto:2005yc},  
inhomogeneities are significantly affected by the additional matter 
couplings, the most sensitive effect being the response of modified gravity to 
spatial variations in the distribution of matter. The situation
bears some resemblance to dark energy models with a coupled dark 
sector\cite{Koivisto:2004ne,Amarzguioui:2004kc,Koivisto:2005nr}. 
In fact, in the conformally equivalent Einstein frame, where the 
line-element (\ref{line}) is given by the metric (\ref{conformal}), 
one finds a scalar field with non-minimal coupling to 
matter\cite{Poplawski:2005an}. 
 
\begin{figure}
\begin{center}
\includegraphics[width=0.47\textwidth]{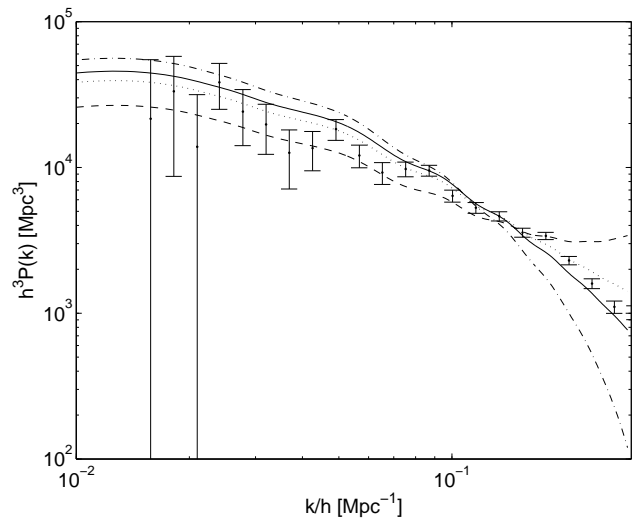}
\caption{\label{mps_pic} The matter power spectra for 
$\beta=-0.00001$ (solid), $\beta=-0.00005$ (dash-dotted),
$\beta=0.00001$ (dotted) and $\beta=0.00005$ (dashed). In all
the cases $\Omega_m = 0.3$ and $n_S=1$. (We do not use the data points 
at the three smallest scales, and instead of the error bars
plotted here we use the exact window functions.)}    
\end{center} 
\end{figure}

We calculate the matter power spectrum,
\be
P(k) = (2\pi k)^{-3}\delta_m^2(k,x=0),
\ee 
where $\delta$ is considered in the Fourier 
space. To fit to the shape of the observed power spectrum, we use the 
best-fit normalization for each model. We assume that the
spectral index lies between $n_S \in [0.8,1.2]$ and marginalize 
over these values. We make a $\chi^2$ fit to the SDSS 
data\footnote{We use the window functions provided at
http://space.mit.edu/home/tegmark/sdss.html.}\cite{Tegmark:2003uf}.
Conservatively, we use only measurements at scales $k < 0.2 h$ Mpc$^{-1}$,
since at smaller scales there are nonlinear effects in the measured 
$P(k)$. We set the Hubble parameter to $h=0.72$ and fix the baryon density 
to $\Omega_b=0.044$ (since especially the latter is 
well determined\cite{Spergel:2003cb}, and now these parameters have
only a slight effect on the initial conditions).

\begin{figure}
\begin{center}
\includegraphics[width=0.47\textwidth]{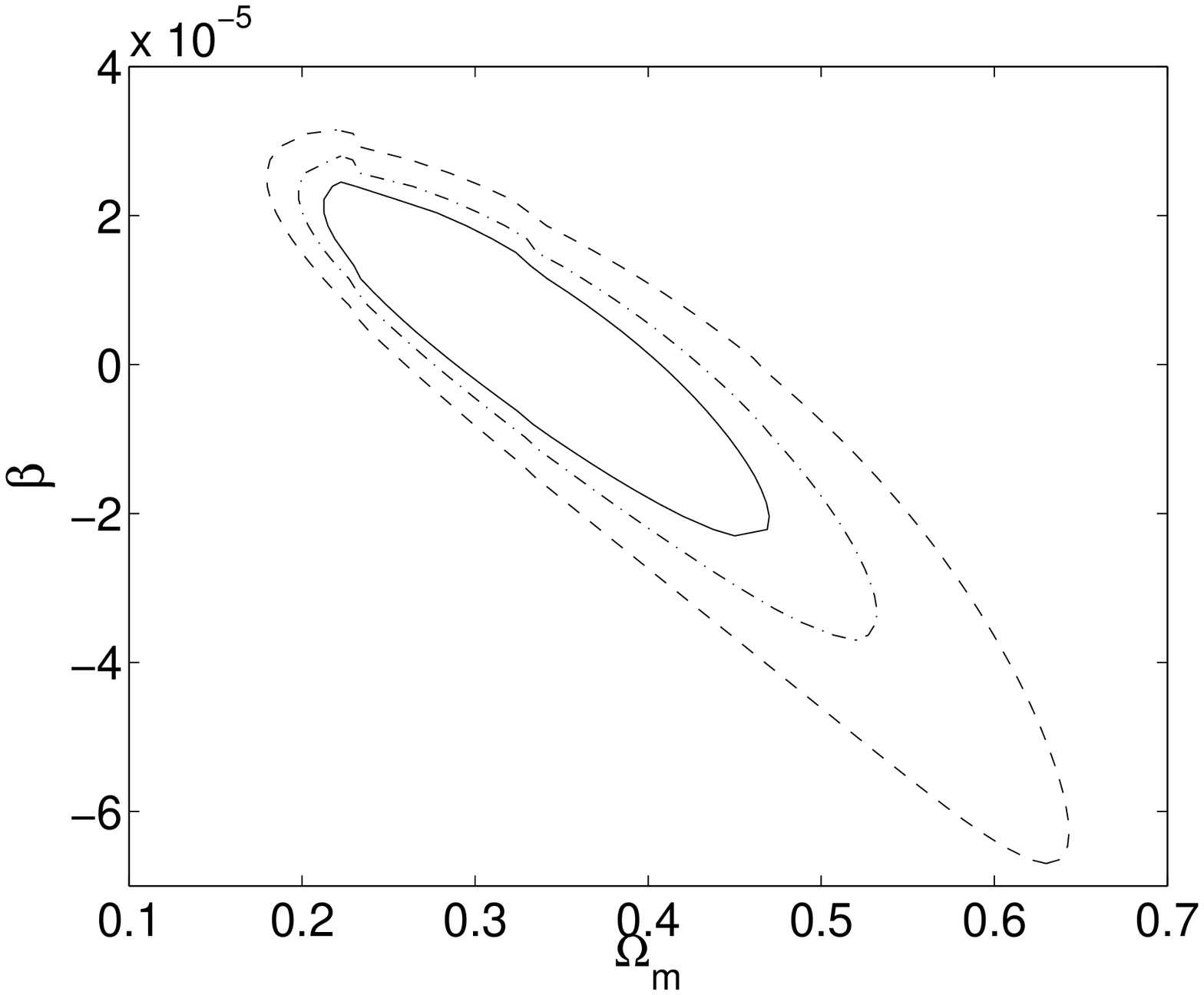}
\includegraphics[width=0.47\textwidth]{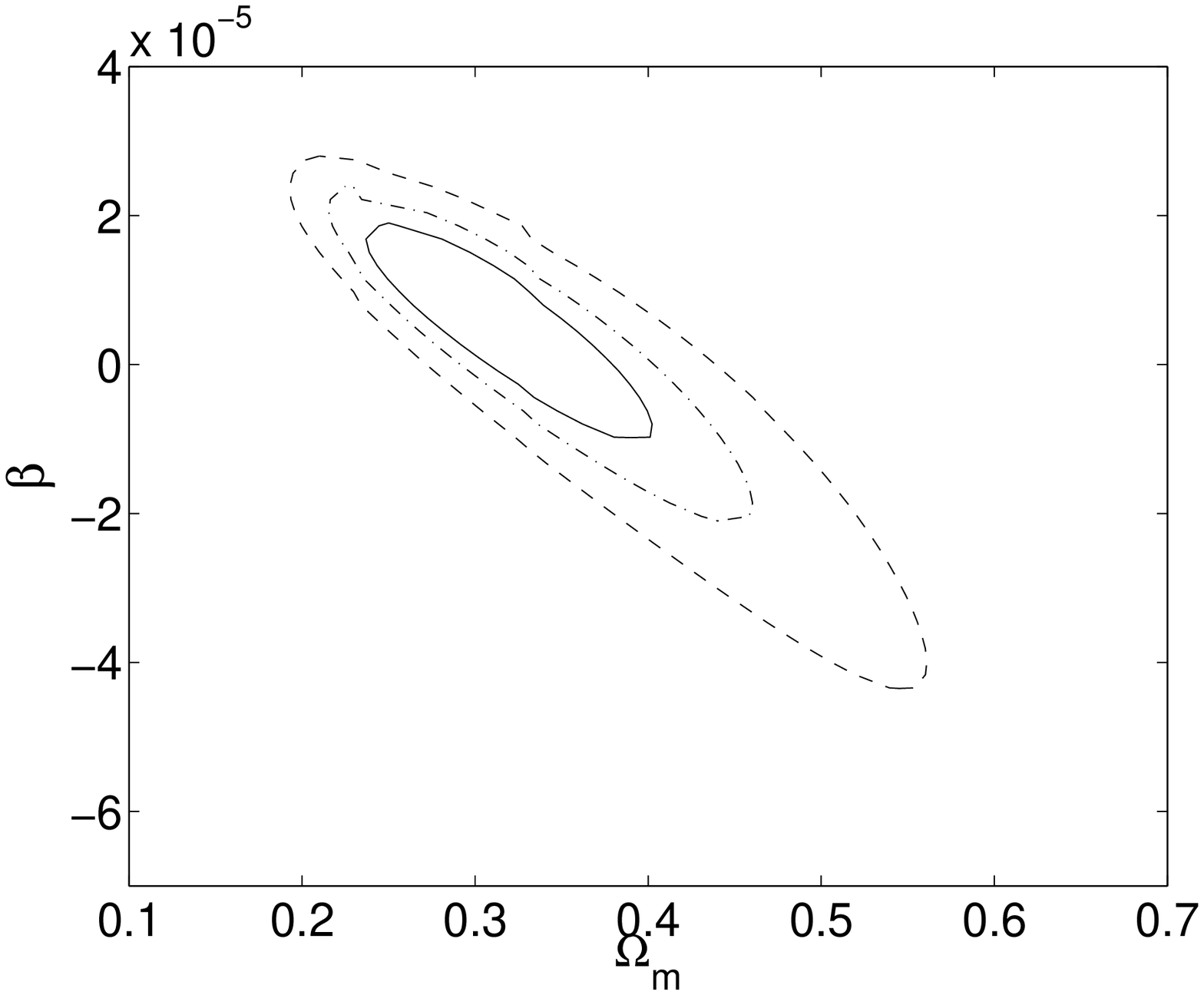}
\caption{\label{mps_limits} The 68, 90 and 99 \% confidence contours 
arising from fitting to the SDSS data. The upper panel is for
$n_S = 1 \pm 0.2$ and the lower panel corresponds to
scale-invariant spectrum of primordial fluctuations.}    
\end{center}
\end{figure}

The resulting power spectra is shown for a few choices of 
the parameter $\beta$ in Fig. \ref{mps_pic}. For these
small $|\beta|$, the gradient term begins to affect perturbations
at just about $k \sim$ 0.1 Mpc$^{-1}$. For much smaller $|\beta|$, the
gradient could be neglected at non-linear scales. For much
larger $|\beta|$, there would be appreciable deviation from 
$\Lambda$CDM cosmology also at very large scales (due to
the $k$-independent coefficients $A$ and $B$ in the evolution equation
Eq.(\ref{deltaevol})), and the effect of the gradient would be large
enough to render these models completely incompatible with 
SDSS data. Thus we can understand why fitting the models to the data 
provides the tight confidence limits plotted in 
Fig. \ref{mps_limits}. For $\beta > 0$, red tilt for the primordial
spectrum is preferred, since the power at small scales is enchanced due to 
modified gravity. Correspondingly, when $\beta < 0$ the best fit
is achieved with blue tilt, $n_S > 1$. The prior we have assumed,
$n_S \in [0.8,1.2]$, is rather loose, since the WMAP experiment is able
to rule out large deviations from scale invariance\cite{Spergel:2003cb}, at 
least when standard assumptions hold in the pre-recombination universe. 
For this reason we plot also, in the lower panel of Fig. \ref{mps_limits}, 
the somewhat tighter constraints which follow from fixing 
the spectral index to $n_S=1$. Marginalizing over $\Omega_m$, we find that 
the favored values of the exponent $\beta$ are smaller than $\sim 3 \cdot 
10^{-5}$ in magnitude. Clearly, inclusion of the smallest scale SDSS data 
points would have tightened these constraints even further. 

\section{Conclusion}

We calculated the matter power spectrum in the Palatini formulation
of modified gravity, and found that the observational constraints
reduce the allowed parameter space to a tiny region around the 
$\Lambda$CDM cosmology. We investigated a specific form for the
gravitational Lagrangian, Eq.(\ref{lagrangian}), but similar
conclusions would probably hold regardless of the parameterization
employed. As already found in\cite{Koivisto:2005yc}, the
problematic features in the perturbation evolution, governed
by Eq.(\ref{deltaevol}), are generic to these models. 
 
It would seem difficult to find a way to produce the observed
large scale structure in the presence of the modified matter
couplings peculiar to these alternative gravity theories when $F$ is 
not very nearly a constant. If
this could be accomplished by invoking exotic properties to
the matter components (isocurvature initial conditions, 
or may be entropic perturbations to cancel the effects of the
couplings), it would not be without adding fine tuning
and ad hoc assumptions to these models. 

Determining consequences of modified gravity at scales of the
Solar system\cite{Olmo:2005zr} or even 
particle physics\cite{Flanagan:2003rb} experiments present 
different challenges as, while exact predictions may be incalculable, 
there can be ambiguities in how to parameterize and interpret the 
possible deviations from Einstein gravity\cite{Vollick:2003ic,Sotiriou:2005xe}. 
Furthermore, it seems that by simple arguments broad classes of models can 
altogether evade the Solar system 
constraints\cite{Sotiriou:2005xe,Cognola:2006eg}.
In contrast, such implications within cosmological perturbation theory as 
discussed here are well understood and robust. Hence it might be 
worthwhile to study to which extent the results obtained here can be 
generalized to other forms of alternative gravity theories. This requires 
detailed work, but would enable probing the possibilities of a more 
fundamental theory of gravitation by the present and up-coming 
high-precision measurements of cosmological large-scale structure. 

\acknowledgments{The author is grateful to H. Kurki-Suonio for valuable 
suggestions on the manuscript, to V. Muhonen for technical advices 
and to the Magnus Ehrnrooth Foundation for support.}

\bibliography{refs}
\end{document}